\documentclass[12pt]{iopart}

\usepackage{iopams}  
\usepackage{amsfonts}
\usepackage{hhline}
\usepackage{braket}
\usepackage{graphicx}
\usepackage{color}
\usepackage{comment}
\begin{document}

\title[Native multiqubit Toffoli gates on ion trap quantum computers]{Native multiqubit Toffoli gates on ion trap quantum computers}

\author{Nilesh Goel}
\address{Department of Physics, Indian Institute of Technology, Hauz Khas, New Delhi, Delhi 110016 India}
\ead{nilesh.goel11@gmail.com}
\author{J. K. Freericks}
\address{Department of Physics, Georgetown University, 37th and O Sts. NW, Washington, DC 20057 USA}
\ead{james.freericks@georgetown.edu}
\vspace{10pt}

\begin{abstract}
We examine the detailed scenario for implementing $n$-control-qubit Toffoli gates and select gates on ion-trap quantum computers, especially those that shuttle ions into interaction zones. We determine expected performance of these gates with realistic parameters for an ion-trap quantum computer and taking into account the time variation of the exchange integrals. This allows us to estimate the errors due to spin-phonon entanglement as well. While there are challenges with implementing these gates, because their performance always has some degree of error, they should be feasible on current hardware, but they may be too slow to be used efficiently in quantum codes on noisy intermediate scale quantum computers.
\end{abstract}

%
%
%
%
%

\section{Introduction}


Quantum computing is usually carried out with a universal gate set that commonly employs single qubit rotations and one two-qubit entangling gate (such as a C-NOT gate). Within this model for computing, programs are written with a small set of single and two qubit gates (often larger than the minimum needed for a universal gate set). Transpiling then converts the gate instructions into the native gate set of the hardware. Often the final set of machine-level instructions is re-optimized by hand to make it run as efficiently as possible on the quantum hardware. 

While this approach is universal and allows the coding to be hardware agnostic, the final step often requires a proper selection of qubits according to their connectivity and a proper reduction of gate operations to the native gates of the hardware. Within this model for computation, more complex gates, like the $n$-bit Toffoli gate, or the select operation (a control from an $n$-qubit ancilla bank that controls on all binary numbers from 0 to $2^n-1$ according to the state of the ancilla), involve complex decompositions into the single and two qubit gates. But, new ion-trap hardware, which allows for shuttling of ions into computational zones, permits one to create native $n$-qubit operations, where $n$ will typically be three or four in near term hardware. 

In a recent paper, Rasmussen \textit{et al.}~\cite{rasmussen} propose a methodology for an $n$-qubit $i$-Toffoli gate, which first nulls the energy of a particular quantum state with a longitudinal magnetic field and then uses a transverse driving field to invert the target qubit if the controls are all in the state with the nulled energy. One can construct a conventional Toffoli gate from this as well, as we describe below. In this work, we provide additional details for how such a native $n$-qubit gate can be implemented within an ion-trap architecture. This requires taking into account a number of issues not discussed in Ref.~\cite{rasmussen} due to how the vibrations are coupled to the effective spin degrees of freedom in the ion-trap quantum computer.


The original M\o llmer-S\o renson gate~\cite{mollmer_sorensen} is a two-qubit entangling gate that is engineered to preserve the number of vibrational quanta excited in the system after the gate operation (but it can only be implemented in the presence of a \textit{longitudinal} magnetic field). This is an important element to have, as the creation of vibrational quanta leads to heating and, ultimately, to decoherence. So, preventing such heating for each two qubit operation is critical to allowing the computer to sequentially carry out a large number of gates with high fidelity and increases the depth of the circuits that can be run on the machine. The generalization by Rassumussen \textit{et al.}~\cite{rasmussen} is in a similar spirit as the original M\o llmer-S\o renson gate, in the sense that it also employs the native Ising coupling that is mediated between the spin degrees of freedom via the spin-dependent laser force (after one integrates out the vibrational degrees of freedom), but it does not provide the same protection that guarantees phonons are not created (although the implementation can be done with phonon creation limited by the detuning of the spin-phonon entangling laser pulses). 


The original ion trap hardware involves ions trapped in a (typically) linear Paul trap with sufficient optical access to allow for lasers that induce the spin-dependent forces and that allow readout of the quantum state via selective atomic cycling transitions (the zero state is bright and the one state is dark, for example). It has been known for years, that one can only hold a modest number of ions within a single linear trap, and still have sufficient control over the system to apply the different quantum gates. To allow these systems to be scaled to large numbers of qubits, the idea of shuttling ions from one region to another was proposed~\cite{shuttling1,shuttling2}. In a shuttling ion-trap computer, one moves ions to different regions of the device in such a fashion that the spin degrees of freedom are not affected by the motion through the system. Within such architectures, one usually has interaction zones, where ions that will be entangled are moved into the zone, while all other ions are far away~\cite{mainz,honeywell}. This allows the quantum computer to have partial reset operations, where the value of a chosen qubit is measured, and if it is a one, it is reset to zero; if it is a zero, it is left unchanged. But, it also allows for $n$-qubit operations to also be employed in this interaction zone, if the zone can hold more than two qubits (holding three or four ions is common). In this work, we rely on such interaction zones, so that "global" operations on all qubits in the interaction zone can be carried out at the same time; we will not need the partial reset capability in this work. Honeywell has a commercial version of such an ion trap computer available~\cite{honeywell}, while they also exist in scientific labs, such as the device in Mainz~\cite{mainz}.


The proposal in Ref.~\cite{rasmussen}, provides a general strategy for implementing the $n$-qubit $i$-Toffoli gate (it is not a standard Toffoli gate, because it flips the target qubit and also provides an additional phase of $i$ to the quantum state when it flips the target). Its emphasis is on employing these gates within error correcting schemes. But such gates arise in many other contexts as well, including time evolution by the sum over unitaries algorithm (where one needs a select operation)~\cite{childs_sum_u} and in the implementation of Kraus maps for driven-dissipative systems (which often use multiple controlled operations)~\cite{childs_dissipative,lorenzo,germans}. They also can be used in simplifying circuits for high-order unitary coupled cluster state preparation in a factorized form, where a double controlled rotation is needed for more efficient implementation of a quad excitation employing two ancilla to reduce the circuit depth~\cite{luogen}.

The organization of this work is as follows: in Sec.~II, we present the formalism for how one describes ion-trap systems; in Sec.~III, we describe the details for how to implement the $i$-Toffoli gate within the ion-trap systems; in Sec.~IV, we present numerical simulations of the gate implementations and we provide conclusions in Sec.~V.

\section{Formalism for an ion-trap quantum computer}

We provide a formal description for how quantum computing is implemented within an ion-trap-based quantum computer. While most of these results here are a backdrop for how to work with ion traps, we are careful to provide explicit details for the time evolution of the system, which involves a number of subtle points.

\subsection{Ion trap effective spin Hamiltonian}


In this work, we choose to be concrete and focus on a specific ion-trap system, as ``the devil is often in the details'' for these systems. We choose to work with the $^{171}$Yb$^+$ ion because it is commonly used in many different quantum systems; it is used in ionQ quantum computers~\cite{ionq} and Honeywell quantum computers~\cite{honeywell}, while calcium is used in the shuttling quantum computers in Mainz~\cite{mainz}. Of course, while the details change for other ions, the general methodology is similar for other systems and it would be too much to try to describe all possible ions here. We follow closely the description given in Ref.~\cite{wang}.


One of the advantages of ion-trap systems is that they work with nearly ideal qubits given by atoms that are employed in atomic clocks---these systems are both well-studied and understood, but they also have coherences for the qubit states that last essentially as long as is needed for any quantum computation. Of course these systems still have significant decoherence, which arises from other mechanisms.

It is common then to use the hyperfine clock states of the ions as the pseudospin states of the quantum computer (these states often have zero total $z$-component of angular momentum, which makes them insensitive to stray magnetic fields, to first-order). The two hyperfine energy levels are the spin up (computational zero) and spin down (computational one) states of the qubit; these states are separated in energy to make it easier to address and measure them. The Hamiltonian that describes the energies of these two levels for an $N$-qubit ion-trap system is then given by 
\begin{equation}
    \hat{\mathcal H}_{0}=\sum_{i=1}^N\frac{\hbar\omega_0}{2}\sigma_j^Z,
    \label{eq:ham_z}
\end{equation}
where $i$ refers to the different ion sites and $\hbar\omega_0$ is the energy difference between the pseudospin states. The spins are described with Pauli matrices, so that $\hat{S}_i^Z=\frac{\hbar}{2}\sigma_i^Z$. In the  $^{171}$Yb$^+$ system, we use the hyperfine clock states with total (nuclear and electronic) $z$-component of angular momentum equal to zero, given by the $\ket{F{=}0,m_F{=}0}$ and $\ket{F{=}1,m_F{=}0}$ states.  The frequency difference between the energy levels is a well-known function of the magnetic field and is given by $\omega_0\approx 12.6$~GHz.

Because the ions are trapped in space (typically at the zeros of the effective electromagnetic pseudopotenial of the system), they  also have vibrational motion relative to these equilibrium positions (we ignore the micromotion in this work). The  phonon Hamiltonian that describes these vibrational modes is expanded in terms of the normal modes of the motion (determined by a simple classical analysis for the pseudopotential)~\cite{james1,james2}
\begin{equation}
    \hat{\mathcal H}_{ph}=\sum_{\alpha=x,y,z}\sum_{\nu=1}^N\hbar\omega_{\alpha\nu}\left (\hat  a^\dagger_{\alpha\nu}\hat a_{\alpha\nu}+\frac{1}{2}\right ),
\end{equation}
where $\nu$ is the index of normal mode and $\alpha$ denotes the spatial direction ($x$, $y$, or $z$). Our notation uses $x,y,z$ to represent the spatial directions of the trap and $X,Y,Z$ to indicate the pseudospin directions. The spatial motion separates into longitudinal ($z$) and transverse ($x$ and $y$)  oscillations. The longitudinal modes are lower in energy than the transverse modes, and one usually works with the transverse modes in an ion trap (where the center-of-mass normal mode is the highest-energy normal mode). Often the trap has some asymmetry in the transverse direction, so the transverse mode frequencies may not always be degenerate as one would find in a perfect trap, but we ignore this small effect here.
Using the raising and lowering operators of the normal modes, we find that the $\alpha^{th}$ spatial component of the position of $i^{th}$ ion is given by $\delta\hat{\textbf{R}}_i^\alpha=\sum_\nu b_i^{\alpha\nu}\sqrt{\frac{\hbar}{2m\omega_{\alpha\nu}}}\left [a_{\alpha\nu}+a_{\alpha\nu}^\dagger\right ]$ where $b_i^{\alpha\nu}$ is the $\nu^{th}$  normal mode at the $i^{th}$ ion site and in the $\alpha$ direction.


In the Yb system, one addresses the ions via a spin-dependent force that is implemented with two laser beams detuned from a Raman transition to a third higher-energy atomic state. By detuning away from the precise transition, one can reduce spontaneous emission effects and continuously drive the qubit as if it is being acted on by a Pauli $X$ operator, which flips the spin from up to down (we will drop calling these clock states pseudospins from now on). A typical experimental implementation uses both red and blue detuned lasers. The effective Hamiltonian for the time-evolution is constructed by first going to the interaction representation with respect to the qubit Hamiltonian in Eq.~(\ref{eq:ham_z}), using the rotating-wave approximation and keeping only the ``slow'' terms, expanding the wavevector dependence in the Lamb-Dicke limit, keeping only the red and blue sideband transitions, and adjusting the phases of the laser beams to an appropriate phase lock (for further details, see Ref.~\cite{wang}). The net effect, is a laser-ion Hamiltonian which drives the spin-flip transition between the qubits and is given by
\begin{equation}
    \hat{\mathcal H}_{LI}(t)=-\hbar\sum_{i=1}^N\Omega_i\Delta\textbf{k}\cdot\delta\hat{\textbf{R}}_i(t)\sigma_i^X \sin(\mu t),
\end{equation}
where $\Omega_i$ is the effective Rabi frequency at the $i^{th}$ ion site and $\mu>0$ is detuning of the laser (also called the beatnote frequency, the phonons are driven to the blue of the transverse center-of-mass mode). Note that in this derivation we dropped a fast oscillating and low amplitude term driven by the carrier transitions.

In addition, we also engineer an additional uniform magnetic field (Zeeman) term by applying two lasers with frequencies separated by $\omega_0$. The phases of the lasers are locked to be out of phase by precisely $\pi$, to get rid of the red and blue sidebands. A second phase can then be adjusted to determine the direction of the effective magnetic field, while an acoustic-optical modulator is employed to adjust the amplitude of the field as desired. This Hamiltonian term is then
\begin{equation}
   \hat{\mathcal H}_B(t)=\sum_{i=1}^N\textbf{B}(t)\cdot\hat{\vec{\sigma}}_i,\label{Zeeman}
\end{equation}
where we use a nonstandard, but common, notation for the sign of the field-spin coupling.
Finally, because we went into the interaction picture with respect to the initial qubit Hamiltonian, the final Hamiltonian for the dynamics of the system is $\hat{\mathcal H}(t)=\hat{\mathcal H}_{ph}+\hat{\mathcal H}_{LI}(t)+\hat{\mathcal H}_B(t).$ Note that all time dependence with respect to $\hat{\mathcal H}_0$ has already been incorporated here, so we do not add additional $e^{\pm i\omega_0 t}$ dependence to the $X$ or $Y$ Pauli spin operators. Our dynamics will now be determined by $\hat{\mathcal H}(t)$.

\subsection{Time-evolution operator}


The time evolution of our system is found from the time-evolution operator. Motivated by the interaction picture, we will develop the time evolution in a hierarchical fashion by factorizing the time-evolution operator in steps. First, we develop the time evolution in the interaction picture with respect to the time-independent piece $\hat{\mathcal H}_{ph}$. This extracts the factor $\exp(-i\hat{\mathcal H}_{ph}t/\hbar)$ to the left, so that 
\begin{equation}
\hat U(t)={\mathcal T}_t e^{-\frac{i}{\hbar}\int_0^t\hat{\mathcal H}(t^\prime)dt^\prime}=e^{-\frac{i}{\hbar}\hat{\mathcal H}_{ph}t}\hat{U}_I(t),
\end{equation}
with $\hat{U}_I(t)$ the standard interaction-picture time-evolution operator with respect to the phonon Hamiltonian. The interaction-picture time-evolution operator satisfies the following equation of motion
\begin{equation}
    i\hbar\frac{d}{d t}\hat{U}_I(t)=\left[e^{\frac{i}{\hbar}\hat{\mathcal H}_{ph}t}\hat{\mathcal H}_{LI}(t)e^{-\frac{i}{\hbar}\hat{\mathcal H}_{ph}t}+\hat{\mathcal H}_B(t)\right]\hat U_I(t),
\end{equation}
because the magnetic field Hamiltonian commutes with the phonon Hamiltonian. Evaluating the Hadamard lemma on the laser-ion Hamiltonian simply changes the
 vibrational raising and lowering operators into the interaction picture operators, given by $a_{\alpha\nu}^I(t)= a_{\alpha\nu}e^{-i\omega_{\alpha\nu}t}$ and $a_{\alpha\nu}^{\dagger I}(t)= a_{\alpha\nu}^\dagger e^{i\omega_{\alpha\nu}t}$. We denote this interaction-picture laser-ion Hamiltonian by $\hat{\mathcal H}_{LI}^I(t)$.


Now, we can perform a second factorization by following the methodology used in Landau and Lifshitz~\cite{LL} and in Gottfried~\cite{gottfried} for the driven harmonic oscillator.  We define an operator $\hat{W}_I(t)=\int_0^tdt'\hat{\mathcal H}_{LI}^I(t')$ that has \textit{no time ordering to it}. Note that $\frac{d}{dt}\hat{W}_I(t)=\hat{\mathcal H}_{LI}^I(t)$. Using this operator we can factorize the unitary operator $\hat{U}_I$ as $\hat{U}_I(t)=e^{-\frac{i}{\hbar}\hat{W}_I(t)}\hat{U}_{II}(t)$. The equation of motion for $\hat{U}_{II}(t)$ becomes
\begin{equation}
     i\hbar\frac{d}{d t}\hat{U}_{II}(t)=e^{\frac{i}{\hbar}\hat{W}_I(t)}\left [-i\hbar\frac{d}{dt}+\hat{\mathcal H}_{LI}^I(t)+\hat{\mathcal{H}}_B(t)\right ]\left \{e^{-\frac{i}{\hbar}\hat{W}_I(t)}\hat{U}_{II}(t)\right \}.
\end{equation}
The Hadamard lemma can be evaluated explicitly for two of the three terms because the commutator of the laser-ion Hamiltonian (in the interaction picture) with itself at two different times produces an operator proportional to $\sigma^X$, which commutes with the laser-ion Hamiltonian. Details are given in Ref.~\cite{wang}. We have
\begin{eqnarray}
e^{\frac{i}{\hbar}\hat{W}_I(t)}\left (i\hbar\frac{d}{dt}\right )e^{-\frac{i}{\hbar}\hat{W}_I(t)}&=&\hat{\mathcal H}_{LI}^I(t)+\frac{i}{2\hbar}[\hat{W}_I(t),\hat{\mathcal H}_{LI}^I(t)],\\
    e^{\frac{i}{\hbar}\hat{W}_I(t)}\hat{\mathcal H}_{LI}^I(t)e^{-\frac{i}{\hbar}\hat{W}_I(t)}&=&\hat{\mathcal H}_{LI}^I(t)+\frac{i}{\hbar}[\hat{W}_I(t),\hat{\mathcal H}_{LI}^I(t)],\\
    e^{\frac{i}{\hbar}\hat{W}_I(t)}\hat{\mathcal{H}}_Be^{-\frac{i}{\hbar}\hat{W}_I(t)}&=&\sum_{i=1}^N\left[\textbf{B}(t)\cdot\hat{\vec{\sigma}}_i+\frac{i}{\hbar}[\hat{W}_I(t),\textbf{B}(t)\cdot\hat{\vec{\sigma}}_i]+\cdots\right];
    \label{IntPicZeemanTerm}
\end{eqnarray}
this last term is an infinite series that typically does not terminate.
For a vanishing transverse magnetic field, given by $B_Y(t)=B_Z(t)=0$, Eq.~(\ref{IntPicZeemanTerm}) simplifies to the Zeeman Hamiltonian of Eq.~(\ref{Zeeman}) even in the interaction picture. In such a case, our interaction picture Hamiltonian reduces to a simple spin Hamiltonian $\hat{\mathcal{H}}_s(t)=\frac{i}{2\hbar}[\hat{W}_I(t),\hat{\mathcal H}_{LI}^I(t)]+\hat{\mathcal H}_B(t)$. Because the spin operators $\sigma_i^X$ in both $\hat{W}_I(t)$ and $\hat{\mathcal H}_{LI}^I(t)$ commute, one can easily derive that the time-dependent Ising spin Hamiltonian is given by
\begin{equation}
    \hat{\mathcal{H}}_s(t)=\sum_{i,j=1}^NJ_{ij}(t)\sigma_i^X\sigma_j^X+\sum_{i=1}^NB_X(t)\hat{{\sigma}}^X_i,
\end{equation}
where we have defined $J_{ij}(t)$ as the time-dependent spin-exchange interaction and it is given by~\cite{monroe_Jij}
\begin{eqnarray}
    J_{ij}(t)&=&\frac{\hbar}{2}\sum_{\nu}\frac{\Omega_i\Omega_j\eta_{x\nu}^2b_i^{x\nu}b_j^{x\nu}}{\mu^2-\omega_{x\nu}^2}\left[\omega_{x\nu}-\omega_{x\nu}\cos2\mu t-2\mu \sin(\omega_{x\nu}t)\sin(\mu t)\right]\label{exchange}\\
    &=&J_{ij}^0+\Delta J_{ij}(t).\label{j}
\end{eqnarray}
where $J_{ij}^0$ is the time-averaged exchange interaction and $\Delta J_{ij}(t)$ is the time-dependent part of the exchange interaction. We introduce the Lamb-Dicke parameter for the $x\nu$ mode, given by $\eta_{x\nu}=\Delta k_x\sqrt{\frac{\hbar}{2m\omega_{x\nu}}}$. $\Delta k_x$ is the $x$-component of the difference in the wavenumber for the two laser beams applied to the ion crystal. Note that the exchange interactions are generically time-dependent. Yet, most analysis of ion-trap simulators assumes the time-independent exchange interactions only. While this may sound like an extreme approximation, it actually works quite well in many cases. Part of the reason why is the phonons in the system tend to dampen the oscillations from the time-dependent spin exchange. Note further that our summation is over all $i$ and $j$. It is common to remove the (unimportant) constant terms that arise when $i=j$ and to combine the terms with unequal indices into a summation over the neighboring pairs, with an exchange constant that is twice as large. In this work, however, we continue to freely sum over both indices, but we set the diagonal terms to zero, because they are unimportant ($J_{ii}^0=0$).

When the transverse field is non-zero (assume here a nonzero field in the Y-direction only), we are left with a residual Hamiltonian given by  $\hat{\mathcal{H}}_{res}(t)=e^{\frac{i}{\hbar}\hat{W}_I(t)}\hat{\mathcal{H}}_{B}(t)e^{-\frac{i}{\hbar}\hat{W}_I(t)}-\hat{\mathcal{H}}_{B}(t)$ after extracting the spin Hamiltonian. This residual Hamiltonian produces the residual spin-phonon interactions and is responsible for additional spin-phonon entanglement (the operator $\hat{W}_I(t)$ also entangles spins with phonons). But, we can still factorize $\hat{U}_{II}$ into two terms: $\hat{U}_{II}(t)=\hat{U}_{spin}(t)\hat{U}_{ent}(t)$, where the spin evolution operator is now given by
\begin{eqnarray}
    \hat{U}_{spin}(t)&=&\mathcal{T}\exp\left[-\frac{i}{\hbar}\int_0^tdt'\left(\sum_{i,j=1}^NJ_{ij}(t')\sigma_i^X\sigma_j^X+B(t')\sum_{i=1}^N\sigma_i^Y\right)\right].\label{spinevol}\
\end{eqnarray}
We defined $\hat{\mathcal{H}}_{spin}(t)=i[\hat{W}_I(t),\hat{\mathcal H}_{LI}^I(t)]/2\hbar+\hat{\mathcal{H}}_{B}(t)$ in the exponent. This factorization gives the following equation of motion for the entanglement evolution operator: $i\hbar\frac{d}{d t}U_{ent}(t)=U_{spin}^\dagger(t)\hat{\mathcal{H}}_{res}(t)U_{spin}(t)U_{ent}(t).$


This entanglement operator is generally quite complicated. If the time dependence of the transverse component of the magnetic field is rapid, then this time-evolution operator is close to one. In other cases, one needs to approximate it, or include it exactly, which can be done by solving for the time evolution of the combined spin and phonon systems. We will not do this here, though.

\section{$n$-qubit $i$-Toffoli gates}


Rasmussen \textit{et al.}~
\cite{rasmussen} describe the basic principles behind creating an $i$-Toffoli gate. We briefly summarize their strategy before describing how to carry it out in an ion-trap system. The methodology uses a technique termed selective subspace inversion. We illustrate the principle using a time-independent Ising Hamiltonian with $n+1$ spins $\hat \mathcal{{H}}_{Is}=\sum_{i, j=1}^{n+1}J_{ij}^0\sigma_i^X\sigma_j^X$ where the exchange coefficients can be thought of as the time-averaged terms in Eq.~(\ref{exchange}). In an actual implementation, we will be working just with the chain of ions in the interaction zone, so we will have $N=n+1$. We represent the states using Ising spins in the $X$-basis and denoted by $|\sigma_1,\sigma_2,\cdots\sigma_{n+1}\rangle$ for the many-body tensor-product state given by $|\sigma_1\rangle\otimes|\sigma_2\rangle\otimes\cdots\otimes|\sigma_{n+1}\rangle$; here $\sigma_i$ is the corresponding eigenvalue in the $X$-direction, which is given by $\pm 1$ (abbreviated to $\pm$). The first spin will be the target qubit; the remaining $n$ spins are the control qubits.  These states are all eigenstates of $\hat \mathcal{{H}}_{Is}$ with vanishing magnetic field, whose energies can be represented by 
\begin{equation}
    E_{|\sigma_1\cdots\sigma_{n+1}\rangle}=\sum_{i,j=1}^{n+1}J_{ij}\sigma_i\sigma_j.
\end{equation}

We define the energy difference between target states, with a fixed control state, via 
\begin{equation}
    \Delta_{\sigma_2,\cdots,\sigma_{n+1}}=E_{|+,\sigma_2,\cdots,\sigma_{n+1}\rangle}-E_{|-,\sigma_2,\cdots,\sigma_{n+1}\rangle}=4\sum_{i=2}^{n+1}J_{i1}^0\sigma_i,\label{Delta}\
\end{equation}
where we used the fact that $J_{ij}^0=J_{ji}^0$. Then the Hamiltonian of the target spin for each control state is given by
\begin{equation}
    \mathcal{H}_{\sigma_1,\cdots\sigma_{n+1}}=\frac{1}{2}\Delta_{\sigma_2,\cdots,\sigma_{n+1}}\sigma_1^X+\frac{1}{2}(E_{|+,\sigma_2\cdots\sigma_{n+1}\rangle}+E_{|-,\sigma_2\cdots\sigma_{n+1}\rangle})\mathbb{I}.
\end{equation}
We think of this situation as the target qubit spin being in an effective magnetic field determined by the control qubits. The goal of the $i$-Toffoli gate is to flip the target spin if the control qubits have a specific value. This can be done by applying a field in the $x$-direction that cancels the control field and also simultaneously applying a field in the $y$-direction that is a $\frac{\pi}{2}$-pulse to flip the spin. Of course, when the longitudinal field is not zeroed out, there is a time evolution of the system, but if the longitudinal field is large for all other cases ($|\Delta-B_x|\gg B_y$), then it is difficult for the target spin to flip in the cases that are not signalled by the control qubits because the system is not on resonance. Hence, the field we apply is 
\begin{equation}
    \vec{\bf B}=-\frac{1}{2}B_x \vec{e}_x+B_y\vec{e}_y,
\end{equation}
and by tuning $B_x$, we will be able to control the gate operation. Our full Hamiltonian is time-independent (except for the turn-on and turn-off of the extra field), so we can immediately determine the time evolution when the extra field is on. We first write the full Hamiltonian using the eigenbases for the control qubits as 
\begin{eqnarray}
\mathcal{H}&=&\sum_{\sigma_2,\cdots,\sigma_{n+1}}\left (\frac{1}{2}(\Delta_{\sigma_2,\cdots,\sigma_{n+1}}-B_x)\sigma_1^X+B_y\sigma_1^Y\right .\\
&+&\left .\frac{1}{2}(E_{|+,\sigma_2,\cdots,\sigma_{n+1}\rangle}+E_{|-,\sigma_2,\cdots,\sigma_{n+1}\rangle})\mathbb{I}_2\right )\otimes |\sigma_2,\cdots,\sigma_{n+1}\rangle\,\langle \sigma_2,\cdots,\sigma_{n+1}|\nonumber
\end{eqnarray}
since the Hamiltonian is diagonal with respect to the control states. The time evolution operator is then straightforward to write down and it is 
\begin{eqnarray}
\hat{U}(t)&=&e^{-i\hat{\mathcal H}t}\label{eq:time}\\&=&\sum_{\sigma_1,\cdots\sigma_{n+1}}e^{-\frac{i}{2}(E_{|+,\sigma_2,\cdots,\sigma_{n+1}\rangle}+E_{|-,\sigma_2,\cdots,\sigma_{n+1}\rangle})t}\nonumber\\
&\times&\Biggr ( \cos\sqrt{{\textstyle\frac{1}{4}}(\Delta_{\sigma_2,\cdots,\sigma_{n+1}}-B_x)^2+B_y^2}t\,|\sigma_1\rangle\,\langle\sigma_1|\nonumber\\
&+&i\frac{\sin\sqrt{\frac{1}{4}(\Delta_{\sigma_2,\cdots,\sigma_{n+1}}-B_x)^2+B_y^2}t}{\sqrt{\frac{1}{4}(\Delta_{\sigma_2,\cdots,\sigma_{n+1}}-B_x)^2+B_y^2}}\Big (\frac{\sigma_1}{2}(\Delta_{\sigma_2,\cdots,\sigma_{n+1}}-B_x)|\sigma_1\rangle\,\langle \sigma_1|\nonumber\\
&~&+B_y\Big (|\sigma_1\rangle\,\langle -\sigma_1|+|-\sigma_1\rangle\,\langle \sigma_1|\Big)\Biggr)\otimes |\sigma_2,\cdots,\sigma_{n+1}\rangle\,\langle\sigma_2,\cdots,\sigma_{n+1}|.\nonumber
\end{eqnarray}
Now it is clear how we proceed. We fix $B_x$ to equal the $\Delta_{\sigma_2,\cdots,\sigma_{n+1}}$ for the desired control qubit sequence and choose the product so that $B_yt=\frac{\pi}{2}$. This then will flip the target qubit and multiply it by a factor of $i$ for the quantum state corresponding to the control qubit; one can also achieve the same result when $B_yt=\frac{(2n+1)\pi}{2}$, but because we want fast operation of the circuits, we will focus on the case with $n=0$). Other control configurations will have a complicated time evolution that is given by Eq.~(\ref{eq:time}). The argument given in Ref.~\cite{rasmussen} is that when $|\Delta-B_x|\gg B_y$, then the phase will be almost that of the free time evolution in the absence of $B_y$. Indeed, by expanding the time evolution when the $B_y$ component is small relative to $|\Delta-B_x|$, the corrections to the proper time evolution (corresponding to $B_y=0$ and given by $\sum_{\sigma_1,\cdots,\sigma_{n+1}}\exp(-iE_{|\sigma_1,\cdots,\sigma_{n+1}\rangle}t)|\sigma_1,\cdots,\sigma_{n+1}\rangle\,\langle\sigma_1,\cdots,\sigma_{n+1}|$) are proportional to $B_y/|\Delta-B_x|=\pi/(2|\Delta-B_x|t)\ll 1$. This is what must be small to guarantee small errors and improve the gate fidelity. But note that the form of this gate is never error free! Nevertheless, it does suggest that running the gate for a longer time with a smaller $B_y$ is preferable to reduce the errors.
This protocol also allows us to run an $n$-qubit ``$i$-select'' gate, which is determined by simply choosing the $B_x$ field appropriately for the desired select operation in addition to the simpler Toffoli gate.

We now describe how such a gate can be used as a standard Toffoli or select gate (without the extra phase of $i$). The strategy is to run the full Toffoli control qubits \textit{and} the target qubit as ancilla. Then, when the target ancilla has been controlled, we can use that ancilla qubit only to control the system, with a standard two-qubit control. We just have to keep in mind the phase kickback of $i$, which may require additional controlled $z$ rotations in any given circuit.


How does this work in an ion trap? While it can be carried out, in principle, for all ion-trap quantum computers, the tight focus required and the ability to null stray light that leaks to other ion positions means it is much more likely to carry this out on an ion-trap quantum computer that uses shuttling of ions. Then one has only the target and control ions in the interaction zone during the application of the gate. So we have $N=n+1$ in this case. By choosing the detuning, one can adjust the values for the exchange coefficients and thereby engineer the different $\Delta$ values. But, these systems have two parity symmetries, which show that when the transverse field is nonzero, the energy eigenstates typically mix two or four parity reflected states. One parity is the spin reflection parity (where the spins in the $X$ and $Z$ directions are inverted, but the $Y$ component is preserved). There is no state in the product-state representation with respect to the $X$ eigenstates that is invariant with respect to this parity operation. The second parity is a spatial reflection symmetry about the center of the ion crystal. There are a number of product states invariant under this parity operation (such as the all up state). The presence of this symmetry need not cause problems with the operation of the gate, but it does mean that the control qubit states no longer are instantaneous eigenstates during the gate operation (due to the nonzero transverse field). Since the gate requires precise timing, it is at these times that the states return to their product state form (valid when $B_y=0$) and thereby timing errors for the gate operation may lead to some undesired entanglement generation in the gate. Or, stated in other words, one must achieve precise timing control to successfully run these gates.


A secondary issue for the gate operation is that the exchange constants are actually time dependent. As has been seen in numerous experiments~\cite{kim,rajibul1,rajibul2}, this time dependence turns out not to have a significant impact on expected results if one approximates instead with the time-averaged exchange constants. This may require the gate operation to be longer than some threshold time for the system to properly average over those time-dependent exchange terms. We examine this question below.

Finally, there are potential issues that arise from the spin-entangling unitary time evolution during the gate operation. While we plan to keep the transverse field small, the product of the field times the time it is applied satisfies $B_yt=\frac{\pi}{2}$ and remains nonnegligible; this could give rise to extra unwanted errors due to the additional entanglement generated by the spin-entanglement piece of the time-ordered product. We will not model such effects in this paper. In general, these effects tend to be small in most calculations, as seen when one simulates the full spin-phonon system. Since the gate involves transverse fields, one cannot ignore the effect of the phonons during the gate operation, unlike the M\o llmer-S\o renson gate, where phonon effects can be completely removed.

\section{Numerical results}


We start our numerical work by considering the $i$-Toffoli gate ($B_x=\Delta_{\sigma_2,\cdots,\sigma_{n+1}}$ when  $\sigma_2,\cdots,\sigma_{n+1}$ are all up). To verify this, we simulate the time-dependent Schrodinger equation for the spin-only Hamiltonian (both with time averaged and time-varying exchange coefficients). To be concrete, we choose to use the experimental parameters used in Ref.~\cite{wang} (and tabulated below). The differential equation is solved by employing the ``ode23'' differential equation solver from MATLAB. 

\subsection{Two-bit $i$-Toffoli gate}

We consider the simplest example of a state selective $i$-Toffoli gate; that is, when $n=2$ (two control qubits, one target qubit), the doubly controlled $i$-Toffoli gate. For this gate, we must have three ions coupled together. Although these ions can be either in the interaction zone of a shuttling ion-trap computer or coupled via tightly focused light on a larger ion-trap quantum computer, we assume the former, so that our entire system consists of just the three ions. We also assume that the laser-spin interaction is given by the form discussed above with realistic parameters for an ion trap. Then, from Eq.~(\ref{Delta}), when all control qubits are in the up state, we have that $\Delta_{\sigma_2,\sigma_{3}}=-4(J_{12}+J_{13})$. We detune our laser frequency close to the transverse center-of-mass phonon mode. then we obtain $J_{12}\approx J_{13}$. Note that $J$ is actually time-dependent [see Eq.~(\ref{exchange})], but as discussed above, it is often appropriate to work with the time-averaged exchange coefficients, so, we set $B_x=-8J_{rms}$, where $J_{rms}=\sqrt{\sum_{i>j}|2J^0_{ij}|^2/N(N-1)}$ where $N=3$ is the total number of ions. 

\begin{figure}[htb]
    \centering
      \includegraphics[width=0.68\textwidth]{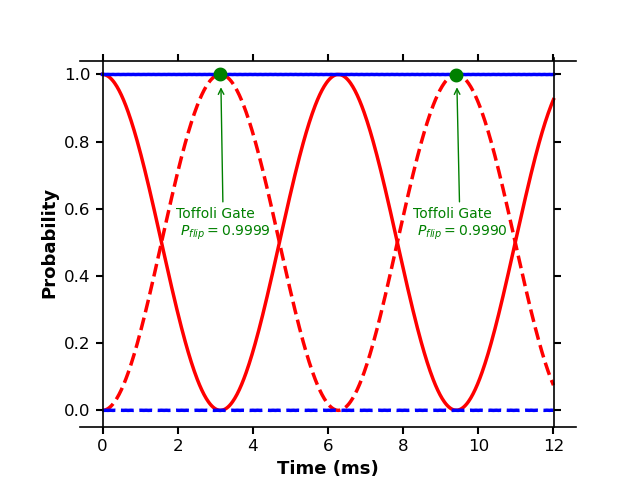}
       \centering
      \caption{Two-$i$-Toffoli gate with static exchange coupling parameters summarized in the text. We plot the probability for the target state to be in $\ket{+}$ (solid line) and $\ket{-}$ (dashed line), when the control qubits are in state $\ket{-,-}$ (red). All the other orientations of control qubits lead to probabilities given by the blue lines. This plot shows near perfect operation when the gate runs over a long operation time of approximately 3~ms. }\label{IdealToff}
    \end{figure}

\begin{figure}[htb]
    \centering
      \includegraphics[width=0.68\textwidth]{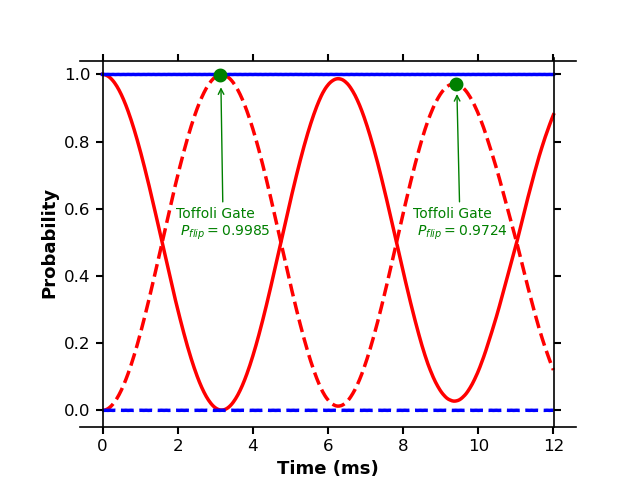}
       \centering
      \caption{Two-$i$-Toffoli gate for time-dependent exchange coupling. We plot the probability for the target state to be in $\ket{+}$ (solid line) and $\ket{-}$ (dashed line) when control qubits are in state $\ket{-,-}$ (red). All the other orientations of control qubits lead to probability variation represented by blue lines. This shows slightly reduced fidelity due to the time-dependent couplings, but the effect is small.}\label{JtToff}
    \end{figure}

We use the following concrete parameters (all frequencies are angular frequencies) $\omega_{CM}=2\pi\times4.63975$~MHz, Lambe-Dicke parameter $\eta=0.06$, Rabi Frequency $\Omega=2\pi\times 369.7$~KHz and a detuning of $\mu=1.0095\omega_{CM}$. The trap anisotropy is $0.2092$ for cases where the total number of ions is even and $0.1$ for cases where the total number of ions is odd (which is what we have here). Since, we have excited near the center-of-mass mode, $J_{ij}$ is approximately the same for all exchange couplings and  $J_{rms}/\hbar=J^0/\hbar\approx2\pi\times 926.019$~Hz. We perform simulations for two cases: (a) We ignore the time-dependence of the exchange interaction and consider the model with static exchange coupling only and (b) we use the exact form of $J(t)$ as given in Eq.~(\ref{j}). For both cases we take $B_y/\hbar=2\pi\times 75.98$~Hz, with case (a) given in Fig.~(\ref{IdealToff}) and case (b) in Fig.~\ref{JtToff}, respectively.

These results clearly show that the best operation occurs for the simple $\pi/2$-pulse. If we extend the simulations to longer times, we find, for example,  that for the $21\pi/2$ flip the probability of success decreased to 0.8298 for time-dependent $J$ and 0.9591 for the static case. This confirms that the gate fidelity continues to drop for longer runs.  

Experience with other simulations that include the full time-dependent effects of phonons showed that when those full effects were included, results were in between those of the static exchange constants and the time-dependent ones. So, to compute what is hopefully a lower bound on the fidelity, we perform the remaining calculations with time-dependent exchange coefficients (but no phonons). 

\begin{figure}[htb]
    \centering
      \includegraphics[width=0.98\textwidth]{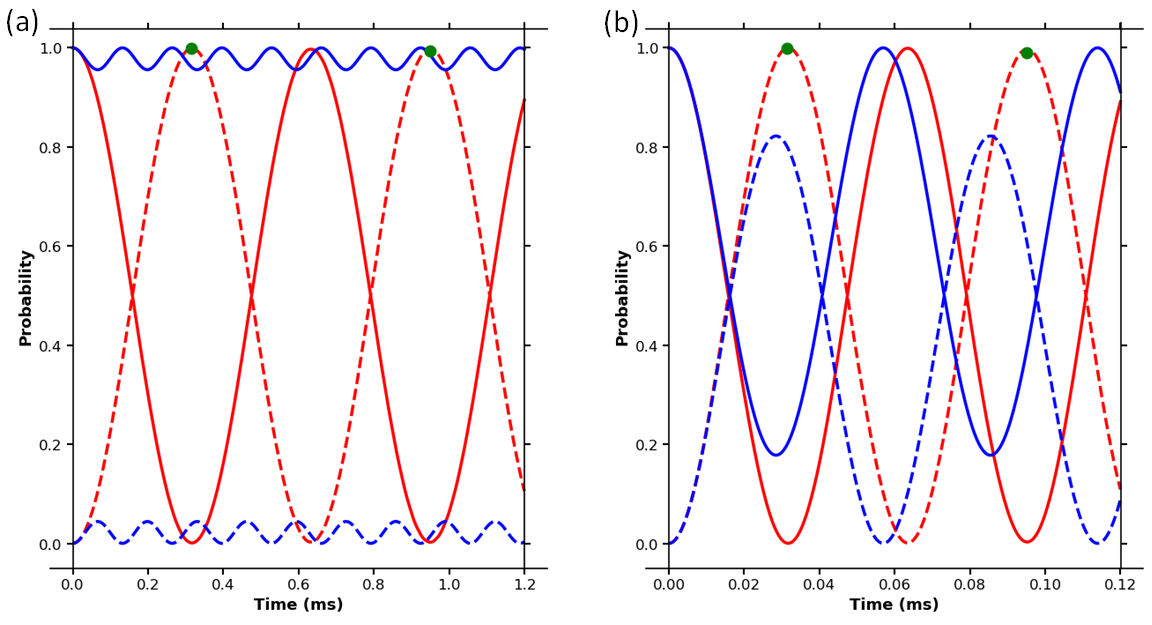}
       \centering
      \caption{Probability for the target qubit to be in the $\ket{+}$ (solid line) and $\ket{-}$ (dashed line) when the control qubits are in the $\ket{-,-}$ (red) and $\ket{+,+}$ (blue) state and $B_y/\hbar$ equal to (a) $2\pi\times 759.8 $~Hz and (b) $2\pi\times 7598$~Hz.}\label{ByDep} 
    \end{figure}

Note that the success of the gate depends on the assumption that $|\Delta-B_x|\gg B_y$ for the off-resonant spaces. We examine how the gate fidelity varies by increasing  $B_y$. This makes the gate run faster, at the expense of reduced fidelity. This can be seen in the Fig.~\ref{ByDep}.

    The probability for the target state to flip, $P_{\scriptsize\textrm{flip}}$, (which is required when control is correct) and to not flip, $P_{\scriptsize\textrm{no flip}}$, (required for all other control combinations) for different values of $B_y$ are listed in the Table \ref{ByDepTable}. One can see that the probability to flip under the control is nearing $0.999$, but the probability to flip for other control bits, is nearly 4\% in the worst case. While this is reasonable operation fidelity, the error is still somewhat high. It may require more engineering of the field, detuning, \textit{etc.}~to achieve better operation.

\begin{table}[]
\centering
\begin{tabular}{ |c|c|c|c|c| } 
 \hline
 \multicolumn{1}{|c|}{Control Qubits} & \multicolumn{2}{|c|}{{$B_y/\hbar=2\pi\times 759.8 $~Hz}} &  \multicolumn{2}{|c|}{$B_y/\hbar=2\pi\times 7598 $~Hz} \\
 \hline
  & $P_{\scriptsize\textrm{flip}}$ & $P_{\scriptsize\textrm{no flip}}$ & $P_{\scriptsize\textrm{flip}}$ & $P_{\scriptsize\textrm{no flip}}$   \\ 
 \hline
 $\ket{-,-}$ & 0.9986 & 0.0014 & 0.9992 & 0.0008 \\
 $\ket{-,+}$ & 0.0373 & 0.9627 & 0.3640 & 0.6360 \\
 $\ket{+,-}$ & 0.0373 & 0.9627 & 0.3640 & 0.6360 \\
 $\ket{+,+}$ & 0.0116 & 0.9884 & 0.5952 & 0.4048  \\
 \hline
\end{tabular}
\caption{Probabilities of flipping the target state for different $B_y$ values and all possible control states in a two-bit $i$-Toffoli gate.}
\label{ByDepTable}
\end{table}

What is interesting about this approach is that the gate fidelity for the target spin to flip when the control is correct, remains essentially unchanged as $B_y$ is increased. The problem we encounter is that faster operation induces target spin flips in cases when the control is such that they should not flip. This will be an unacceptable gate error, as the probabilities for these ``false'' flips are sizable. We find a good transverse field seems to be equal to be about 1/10th of our $J_{rms}$.  Note, however, that because the periods of the oscillations of the target qubit probabilities depend on the different control bit states, it may be possible to engineer a fast gate operation where we have high fidelity for all possible control qubit configurations by engineering the highest fidelity for the other control bit configurations to occur at the sames time as the $\pi/2$ pulse for the all down control. This requires a significant optimization procedure, which is not worthwhile to attempt without having the specific parameters of operation for a specific quantum computer hardware.

\begin{figure}[htb]
    \centering
      \includegraphics[width=\textwidth]{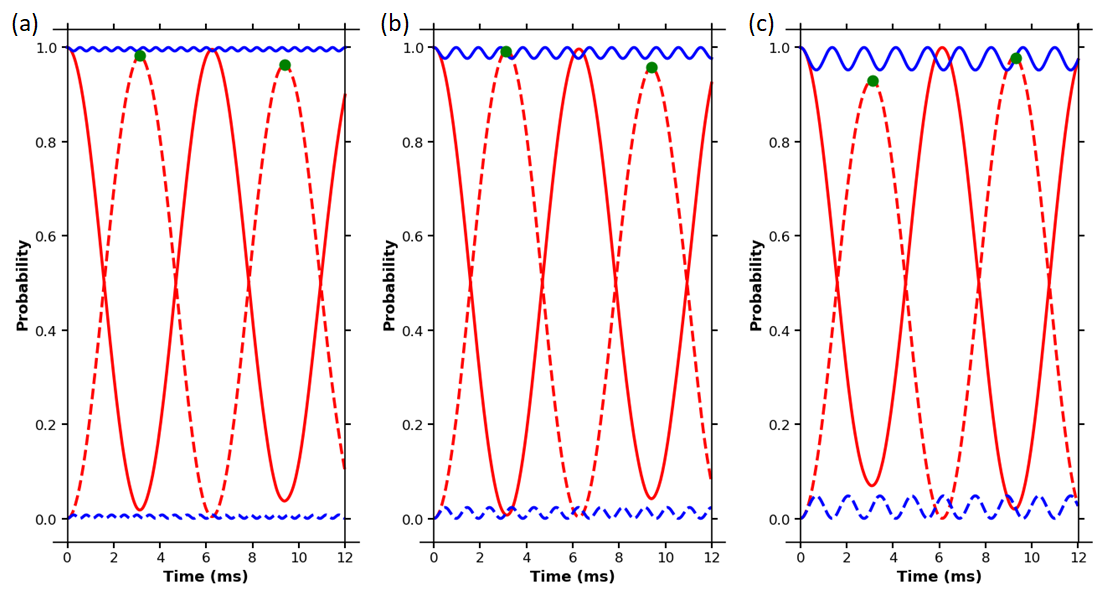}
       \centering
      \caption{Probability for the target qubit to be in the state $\ket{+}$ (solid line) and $\ket{-}$ (dashed line) when the control qubits are in the $\ket{-,-}$ (red) and $\ket{+,+}$ (blue) states for $B_y/\hbar=2\pi\times 75.98$~Hz and the detuning $\mu/\omega_{CM}$ equal to (a) 1.0380, (b) 1.0655, (c) 1.0950.}\label{muDep}
    \end{figure}

Next, we examine the effect of the detuning, as we move further from the transverse center-of-mass mode frequency (moving further into the blue). For the case where $B_y/\hbar=2\pi\times 75.98$~Hz, we find a degradation of the probability to not flip when the control qubits are in the wrong state, and also a degradation of the switching probability, when the control configuration is correct. These results are shown in Fig.~\ref{muDep}. 

    The most likely reason for this is that the energy spacings decrease as the detuning is made larger, which creates a higher probability for an error with respect to the state of the control qubits.
    
    \subsection{n-bit $i$-Toffoli gate}
    
\begin{figure}[htb]
    \centering
      \includegraphics[width=\textwidth]{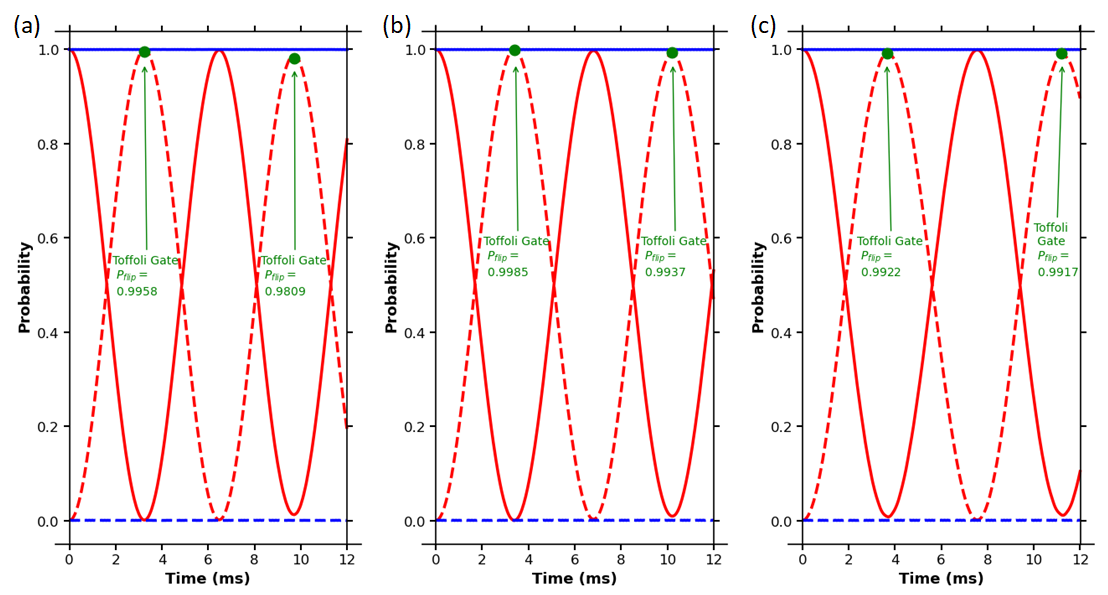}
       \centering
      \caption{Probability for the target qubit to be in the state $\ket{+}$ (solid line) and $\ket{-}$ (dashed line) when the control qubits are in the all down (red) and all up (blue) states for $n$ equal to (a) 3, (b) 4 and (c) 5.}\label{nDep}
    \end{figure}

    Since we have proposed a method to implement an $n$-bit $i$-Toffoli gate, we also ran simulations for $n=3$, $4$ and $5$ by taking $\mu$ values such that $J_{rms}$ was nearly equal for all cases. These simulations are shown in Fig.~\ref{nDep}. The detunings used were:
    
    \begin{itemize}
        \item n=3: $\mu=1.00713\omega_{CM}$ $\Rightarrow$ $J_{rms}/\hbar=2\pi\times 926.307$~Hz
        \item n=4: $\mu=1.00571\omega_{CM}$ $\Rightarrow$ $J_{rms}/\hbar=2\pi\times 925.924$~Hz
        \item n=5: $\mu=1.00476\omega_{CM}$ $\Rightarrow$ $J_{rms}/\hbar=2\pi\times 925.876$~Hz
    \end{itemize}
    
    For these parameters, the probabilities to flip for different combinations of control qubits for n=3, 4 and 5 have been tabulated in Tables $\ref{3Bit}$, $\ref{4Bit}$ and $\ref{5Bit}$. We see from the tables that probability to flip is large only when control is correct; i.e., $\ket{-,-,-}$. 
    
    One can see that the fidelity remains quite high for these cases. This means scaling these gates to larger $n$ values should be viable if the gate operation is short enough for efficient use in digital circuits.

\begin{table}[]
\centering
\scriptsize
\begin{tabular}{ |c|c|c|c|c|c|c|c|c| } 
 \hline
  Control & $\ket{-,-,-}$ & $\ket{-,-,+}$  & $\ket{-,+,-}$ & $\ket{-,+,+}$ & $\ket{+,-,-}$ & $\ket{+,-,+}$  & $\ket{+,+,-}$ & $\ket{+,+,+}$  \\ 
 \hline
 $P_{\scriptsize\textrm{flip}}$ & 0.99582 & 0.00029 & 0.00029 & 0.00004 & 0.00029 & 0.00004 & 0.00004 & 0.00003  \\
 \hline
\end{tabular}
\normalsize
\caption{Probability to flip the target state for all possible control states of a 3-bit $i$-Toffoli gate. Note that the probability not to flip is $1-P_{\scriptsize\textrm{flip}}$.}
\label{3Bit}
\end{table}

\begin{table}[]
\centering
\begin{tabular}{ |c|c|c|c|c| } 
 \hline
  Control & $\ket{-,-,-,-}$ & $\ket{-,-,-,+}$  & $\ket{-,-,+,-}$ & $\ket{-,-,+,+}$ \\ 
 \hline
 $P_{\scriptsize\textrm{flip}}$ & 0.99851 & 0.00006 & 0.00006 & 0.00004  \\
 \hhline{|=|=|=|=|=|}
  Control & $\ket{-,+,-,-}$ & $\ket{-,+,-,+}$  & $\ket{-,+,+,-}$ & $\ket{-,+,+,+}$ \\ 
 \hline
 $P_{\scriptsize\textrm{flip}}$ & 0.00005 & 0.00004 & 0.00004 & 0.00005  \\
 \hhline{|=|=|=|=|=|}
  Control & $\ket{+,-,-,-}$ & $\ket{+,-,-,+}$  & $\ket{+,-,+,-}$ & $\ket{+,-,+,+}$ \\ 
 \hline
 $P_{\scriptsize\textrm{flip}}$ & 0.00005 & 0.00004 & 0.00004 & 0.00005  \\
 \hhline{|=|=|=|=|=|}
  Control & $\ket{+,+,-,-}$ & $\ket{+,+,-,+}$  & $\ket{+,+,+,-}$ & $\ket{+,+,+,+}$ \\ 
 \hline
 $P_{\scriptsize\textrm{flip}}$ & 0.00004 & 0.00003 & 0.00003 & 0.00009  \\
 \hline
 
\end{tabular}
\caption{Probability to flip the target state for all possible control states of a 4-bit $i$-Toffoli gate. Note that the probability not to flip is $1-P_{\scriptsize\textrm{flip}}$.}
\label{4Bit}
\end{table}

\begin{table}[]
\centering
\begin{tabular}{ |c|c|c|c|c| } 
 \hline
  Control & $\ket{-,-,-,-,-}$ & $\ket{-,-,-,-,+}$  & $\ket{-,-,-,+,-}$ & $\ket{-,-,-,+,+}$ \\ 
 \hline
 $P_{\scriptsize\textrm{flip}}$ & 0.99223 & 0.00027 & 0.00025 & 0.00013  \\
 \hhline{|=|=|=|=|=|}
 
   Control & $\ket{-,-,+,-,-}$ & $\ket{-,-,+,-,+}$  & $\ket{-,-,+,+,-}$ & $\ket{-,-,+,+,+}$ \\ 
 \hline
 $P_{\scriptsize\textrm{flip}}$ & 0.00025 & 0.00013 & 0.00015 & 0.00979  \\
 \hhline{|=|=|=|=|=|}
 
   Control & $\ket{-,+,-,-,-}$ & $\ket{-,+,-,-,+}$  & $\ket{-,+,-,+,-}$ & $\ket{-,+,-,+,+}$ \\ 
 \hline
 $P_{\scriptsize\textrm{flip}}$ & 0.00025 & 0.00019 & 0.00019 & 0.00898  \\
 \hhline{|=|=|=|=|=|}
 
   Control & $\ket{-,+,+,-,-}$ & $\ket{-,+,+,-,+}$  & $\ket{-,+,+,+,-}$ & $\ket{-,+,+,+,+}$ \\ 
 \hline
 $P_{\scriptsize\textrm{flip}}$ & 0.00019 & 0.00898 & 0.00898 & 0.00002  \\
 \hhline{|=|=|=|=|=|}
 
   Control & $\ket{+,-,-,-,-}$ & $\ket{+,-,-,-,+}$  & $\ket{+,-,-,+,-}$ & $\ket{+,-,-,+,+}$ \\ 
 \hline
 $P_{\scriptsize\textrm{flip}}$ & 0.00025 & 0.00019 & 0.00019 & 0.00898  \\
 \hhline{|=|=|=|=|=|}
 
   Control & $\ket{+,-,+,-,-}$ & $\ket{+,-,+,-,+}$  & $\ket{+,-,+,+,-}$ & $\ket{+,-,+,+,+}$ \\ 
 \hline
 $P_{\scriptsize\textrm{flip}}$ & 0.00019 & 0.00898 & 0.00898 & 0.00002  \\
 \hhline{|=|=|=|=|=|}
 
   Control & $\ket{+,+,-,-,-}$ & $\ket{+,+,-,-,+}$  & $\ket{+,+,-,+,-}$ & $\ket{+,+,-,+,+}$ \\ 
 \hline
 $P_{\scriptsize\textrm{flip}}$ & 0.00010 & 0.01366 & 0.01030 & 0.00001  \\
 \hhline{|=|=|=|=|=|}
 
   Control & $\ket{+,+,+,-,-}$ & $\ket{+,+,+,-,+}$  & $\ket{+,+,+,+,-}$ & $\ket{+,+,+,+,+}$ \\ 
 \hline
 $P_{\scriptsize\textrm{flip}}$ & 0.01030 & 0.00001 & 0.00002 & 0.00001  \\

 \hline
 
\end{tabular}
\caption{Probability to flip the target state for all possible control states of a 5-bit $i$-Toffoli gate. Note that the probability not to flip is $1-P_{\scriptsize\textrm{flip}}$.}
\label{5Bit}
\end{table}

\subsection{Two-bit $i$-select gate}    

A select gate is a more complicated gate than a Toffoli gate. In a select gate, the target will flip for any of the chosen controls and we have a tunable (or programmable) ability to select the control configuration (in this case by adjusting the $B_x$ value). The select gate is used for the sum over unitaries algorithm and is likely to have wide application in fault-tolerant quantum computers. Here, we discuss the possible applications in noisy intermediate scale quantum computers.

In the previous $i$-Toffoli gates described above, we used the center-of-mass mode because the longitudinal magnetic field lifts the degeneracy of the control state and detuning to the blue of the transverse modes produces no cross-talk with other phonon modes, since they are all lower in energy. But, when we need to also perform state inversion for control qubits in the $\ket{-,+}$ state, we find that  $\Delta_{\sigma_2,\sigma_{3}}$ is the same for both control configurations $\ket{-,+}$ and $\ket{+,-}$ in this case. Hence, we cannot invert the target qubit for each control-qubit configuration separately. To achieve a true select qubit operation, we can use detuning from the zigzag (ZZ) mode of the 3-ion system, which has $J_{12}=-2J=J_{23}$ and $J_{31}=J$ where $J$ is given by Eq.~(\ref{j}) with the phonon eigenvector given by $b^{ZZ}=\{0.4082, -0.8164, 0.4082\}$. In this mode, if we assign site $1$ as the target ion site, we obtain four different values of $\Delta_{\sigma_2,\sigma_{3}}$ for the four different combinations of control qubits. 

\begin{figure}[htb]
    \centering
      \includegraphics[width=0.9\textwidth]{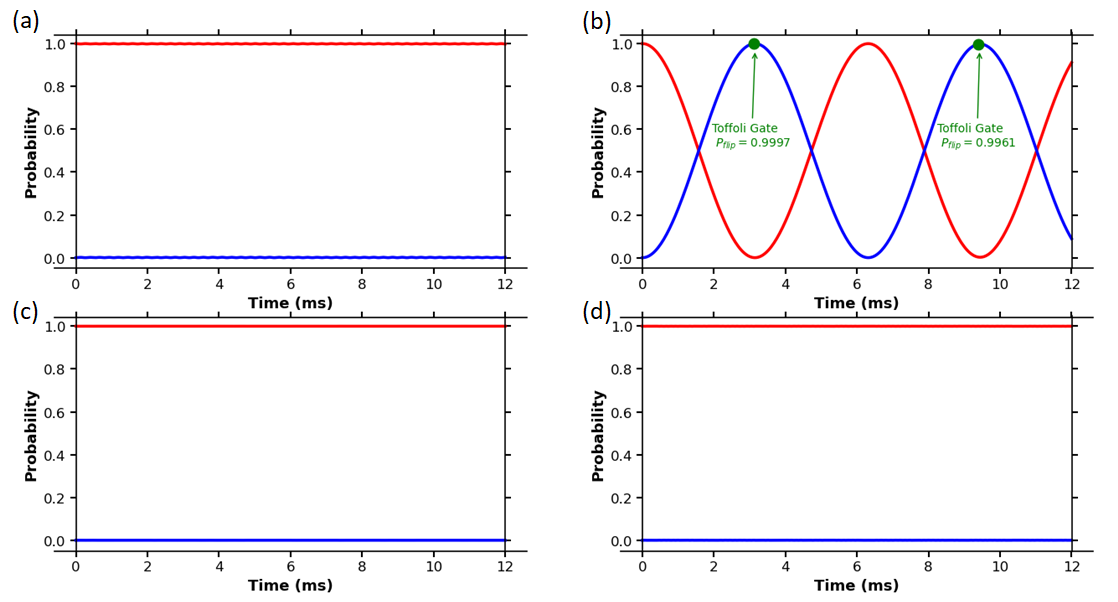}
       \centering
      \caption{Probability of the target state being in $\ket{+}$ (red) and $\ket{-}$ (blue) when $B_x=12J$ and the control qubits are in state (a) $\ket{-,-}$, (b) $\ket{-,+}$, (c) $\ket{+,-}$ and (d) $\ket{+,+}$. Other controlled cases are similar to this and are not shown. Detailed quantitative data can be found in the accompanying table. }\label{Selectmp}
    \end{figure}

We obtain $\Delta\approx 4J$, $12J$, $-12J$ and $-4J$ for control qubit state $\ket{-,-}$, $\ket{-,+}$, $\ket{+,-}$ and $\ket{+,+}$, respectively. Therefore, we can simply adjust our longitudinal magnetic field according to whichever subspace inversion is required and we have created an $i$-select gate. To use the ZZ mode, we used the fact that the transverse phonon mode frequency for zigzag mode is 0.9879 of the CM mode frequency. Thus, $\omega_{ZZ}=2\pi\times4.58361$~MHz. We used the detuning of $0.9905\omega_{ZZ}$ (to the red of the zigzag mode) to perform the simulations of the select gate. The probabilities of flipping the target state for a given subspace are tabulated in Table \ref{SelectGate}. The action of select gate for the subspace $\Delta=12J$ has also been plotted in Fig.~\ref{Selectmp} in which we can see that the target state flips only when control qubits are in $\ket{-,+}$ state. One can see the performance here is excellent, even better than the center-of-mass tuned $i$-Toffoli gates. Of course one does have to be careful when detuning to the red of the transverse zigzag mode not to simultaneously be detuned close to the longitudinal modes, but that does not occur in this case.
The longitudinal mode frequency of the zigzag mode is given by $\omega_{ZZ,L}=2.408\omega_{CM,L}$. Since, for $n=2$; i.e., a 3-ion system, our anisotropy is 0.1, we have that $\omega_{CM,L}=0.1\omega_{CM}$ which implies that $\omega_{ZZ,L}=0.2408\omega_{CM}$. This is far from the red-detuned transverse zigzag mode at $0.9879\omega_{CM}$. This means we are unlikely to excite longitudinal modes when we detune to the red of the transverse zig-zag mode.

\begin{table}[]
\centering
\begin{tabular}{ |c|c|c|c|c| } 
 \hline
 \multicolumn{1}{|c|}{} & \multicolumn{4}{|c|}{{$B_x$ in terms of $J$}} \\
 \hline
  Control Qubits & $\approx 4J$ & $\approx 12J$ & $\approx -12J$ & $\approx -4J$   \\ 
 \hline
 $\ket{-,-}$ & 0.9995 & 0.0005 & 0.0004 & 0.0002 \\
 $\ket{-,+}$ & 0.0007 & 0.9997 & 0.0002 & 0.0004 \\
 $\ket{+,-}$ & 0.0003 & 0.0001 & 0.9945 & 0.0006 \\
 $\ket{+,+}$ & 0.0006 & 0.0004 & 0.0007 & 0.9972  \\
 \hline
\end{tabular}
\caption{Probabilities of flipping the target state for different $B_x$ and control states in a 2-bit select gate at $B_yt=\pi/2$. Note that the probability not to flip is $1-P_{\scriptsize\textrm{flip}}$. This has better than 0.99 performance and is approaching 0.999 in many cases.}
\label{SelectGate}
\end{table}

\section{Conclusions}

In this work, we examined the possibility of executing Toffoli and select gates as multi-qubit gates in ion-trap-based quantum computers. We find that generically, one can achieve fairly good fidelity of operation, but the gates tend to be slow (about an order of magnitude or more slower than two-qubit entangling gates). It is possible that one can improve the speed of operation by precise tuning of parameters in the time-evolving effective Hamiltonian of the spin system. We did not try to engage in this engineering here, because it requires detailed knowledge of the parameters in the specific quantum computer. 

Our Toffoli gates were implemented by detuning near the center-of-mass mode and the select gate by detuning near the zig-zag mode. The latter tended to have better performance, even though that detuning has the possibilitiy to drive other phonon modes more easily.

We did not include the phonon creation in these calculations. Driving close to phonon mode can create phonons and hence heat the system. This is a concern that should be addressed in future work as well, if one tries to implement these types of gates in actual harware.

In summary, this work shows that it should be possible to implement these types of multi-qubit gates, although achieving high gate fidelity will be a challenge that faces a number of additional obstacles not needing to be dealt with for two qubit entangling gates. We hope that future work will investigate this behavior on quantum hardware, possibly leading to the introduction of these gates as part of the native gate set of ion-trap quantum computers.


\ack

This work was supported by the National Science Foundation under grant number
PHY-1915130. In addition, JKF was supported by the McDevitt bequest at Georgetown

~\\
{\bf \large References}
~\\

\end{document}